\documentclass[prl,twocolumn,amsmath,amssymb,showpacs]{revtex4}
\usepackage{graphicx}
\begin{document}
\title{Nonlocal correlations in normal-metal superconducting systems}

\author{P. Cadden-Zimansky and V. Chandrasekhar}

\affiliation{Department of Physics and Astronomy, Northwestern
University, Evanston, IL 60208}

\date{\today}

\begin{abstract}
We examine nonlocal effects between normal-metal gold probes connected by superconducting aluminum.  For highly transparent Au/Al interfaces, we find nonlocal voltages that obey a spatial and temperature evolution distinct from the nonequilibrium charge imbalance signals usually found in such systems.  These voltages are consistent with the predicted effects of crossed Andreev reflection and elastic cotunneling, effects that involve coherent correlations between spatially separated electrons.
\end{abstract}
\pacs{74.45.+c, 03.67.Mn, 74.78.Na}
\maketitle

Nonlocal, coherent effects are hallmarks of quantum
mechanics.  Recently, interest has grown in nonlocal
processes that coherently couple electrons in spatially
separated normal metals (N) linked by a superconductor (S).\cite{byers,deutscher,falci,bignon}  The correlations induced between the spatially separated electrons are predicted to give rise to nonlocal voltages in response to a drive current below the superconducting transition.  Here
we present experimental evidence for such nonlocal voltages in normal-superconducting
systems with highly transparent interfaces, and demonstrate that they exhibit the predicted spatial evolution.

\begin{figure}[!]
      \includegraphics[width=4.5cm]{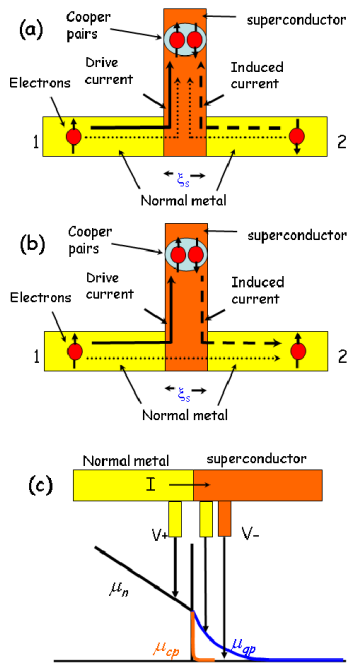}
      \caption{{\small (color online) Schematic representation of electronic processes that contribute to nonlocal voltages in normal-metal/superconductor structures.  \textbf{(a)}  In crossed Andreev reflection (CAR), an electron from one normal metal lead (1) combines with an electron of opposite spin orientation from the second normal lead (2) to form a Cooper pair in the superconductor.  \textbf{(b)}  In elastic cotunneling (EC), a electron tunnels from normal lead 1 to normal lead 2 through the effective tunnel barrier represented by the superconducting gap.  \textbf{(c)}  In charge imbalance (CI), a quasiparticle current injected from the normal metal into the superconductor results in nonequilibrium electrochemical potentials $\mu_{qp}$ and $\mu_{cp}$ (for quasiparticles and Cooper pairs respectively) that relax to the electrochemical potential in the bulk of the superconductor on different length scales.  The schematic depicts local CI; a nonlocal CI contribution also exists in regions of the superconductor near the NS interface not in the current path.\vspace{-.7cm}} \label{fig1}}
\end{figure}

Figure \ref{fig1}(a) and (b) illustrate two processes that are predicted to coherently couple electrons in normal metals connected to a common superconductor.  In crossed Andreev reflection (CAR), electrons from one normal metal are coupled to electrons in the other normal metal through a nonlocal analog of Andreev reflection, so long as the width of the superconductor between the two normal metals is comparable to or less than the superconducting coherence length, $\xi_S$.  In elastic co-tunneling (EC), electrons with predominantly sub-gap energies tunnel directly from one normal metal to another.  In both cases, the electrons remain in a coherent entangled state, coupled through their interaction with the superconductor.\cite{deutscher}  For CAR, a current driven from one normal metal into the superconductor induces a current from the second normal metal into the superconductor due to the coupling of pairs of electrons in the two N leads; for EC, the direction of the induced current is opposite.  In both CAR and EC, the width of the superconductor between lead 1 and 2 must be comparable to the superconducting coherence length, $\xi_S$.

Experimentally, it is more convenient  to measure the voltage on the second lead with respect to the superconductor:  for CAR, the drive current will induce a negative voltage on the second N lead, while for EC, the voltage is positive.  The voltage is nonlocal in the sense that no net current flows through the second NS interface or the second N lead in the measurement.  Since CAR and EC depend on the spin orientation of the electrons, both processes are affected if there is a finite spin polarization in the normal leads, such as would be the case if the normal leads were ferromagnetic (F).  For example, in the extreme limit of half-metallic ferromagnets where only one spin orientation is present in each ferromagnetic lead, EC is favored and CAR is completely suppressed if the magnetizations of the two ferromagnets are parallel, while CAR is favored and EC completely suppressed if the magnetizations are antiparallel.   For extended NS contacts the nonlocal voltages arising from EC and CAR are predicted to decay exponentially with the width of the superconductor, $L$, over the length scale of $\xi_S$.\cite{feinberg}

A third process may complicate the observation of CAR and EC in NS and FS structures.  This is the phenomenon of charge imbalance (CI).\cite{clarke,tinkham,schmid}  The physics of this process is shown schematically in Fig. \ref{fig1}(c).   A current, $I$, of quasiparticles injected from the normal metal gives rise to a nonequilibrium distribution of quasiparticles and Cooper pairs in the superconductor near the NS interface, which in turn leads to different electrochemical potentials $\mu_{qp}$ and $\mu_{cp}$ near the NS interface for the quasiparticles and Cooper pairs.  $\mu_{qp}$ and $\mu_{cp}$ can be probed by normal metal and superconducting probes respectively.
$\mu_{cp}$ relaxes to the electrochemical potential in the bulk of the superconductor $\mu_0$ over the length scale $\xi_S$, which is typically of the order of 100 nm for conventional superconductors.  $\mu_{qp}$, on the other hand, relaxes to $\mu_0$ over the charge imbalance length $\Lambda_{Q^*}$, which can be much longer.\cite{chi, stuivinga}  If the potential at the interface is taken to be zero, $\mu_{qp}$ in the superconductor has the spatial dependence\cite{chien}
\begin{equation}
\mu_{qp}(x)= e \Lambda_{Q^*} \rho_S I \tanh(x/\Lambda_{Q^*})
\label{eqn1}
\end{equation}
where $x$ is the distance from the interface and $\rho_S$ is the resistance per unit length of the superconducting wire in its normal state.  

From the equation above, the asymptotic value of $\mu_{qp}$ deep in the superconductor is $e \Lambda_{Q^*} \rho_S I=\mu_0$.
A superconducting lead used to measure the voltage in the superconductor near the NS interface will probe $\mu_{cp}$; beyond a distance $\xi_S$ from the interface, this is essentially $\mu_0$.  
On the other hand, a normal voltage lead placed on the superconductor will measure $\mu_{qp}$.  In this case, there will be a difference $\mu_{qp}(x)-\mu_0$ between the voltage measured in the superconductor by the normal lead and any superconducting lead located more than a distance $\xi_S$ from the interface.  The resulting enhancement in the resistance will decay with the distance $x$ from the interface as $\Delta R(x)=\rho_S \Lambda_{Q^*}(1- \tanh(x/\Lambda_{Q^*}))$.  

$\Lambda_{Q^*}$ is expected to diverge at temperatures near the transition temperature $T_c$ of the superconductor or at quasiparticle energies near the superconducting gap.\cite{schmid}  The CI contribution vanishes in the limit of low temperatures and low energies, as shown by a number of experiments over the last thirty years.\cite{chi,hsiang,stuivinga,van harlingen,mamin,strunk,chien}  In these experiments, however, the induced charge imbalance was \textit{local}, in the sense that the nonequilibrium distributions were probed along the current path, as depicted in Fig. \ref{fig1}(c).  Intuitively, one would also expect a \textit{nonlocal} CI contribution in parts of the superconductor near the NS interface but outside the current path, although the effective $\Lambda_{Q^*}$ might be different.  A detailed theoretical description of nonlocal CI is yet to be worked out.

Two groups have recently reported observations of CAR and EC in patterned metallic heterostructures.  Beckmann  \textit{et al.}\cite{beckmann}  measured the nonlocal resistance of a device consisting of multiple ferromagnetic Fe wires in contact with an Al wire, where the magnetization of the Fe wires could be aligned parallel or antiparallel by means of an external magnetic field.  They observed a very small difference $\Delta R$ in the nonlocal resistance between the parallel and antiparallel configurations below $T_c$ of the Al wire, consistent with that expected for CAR and EC.  This signal was superposed on a CI contribution that was almost three orders of magnitude larger.  Russo \textit{et al.}\cite{russo} measured the nonlocal resistance of NS structures with tunneling interfaces as a function of the voltage bias.  They observed a positive nonlocal voltage at zero bias that turned into a negative nonlocal voltage with increasing voltage bias, which they ascribed to a transition from EC to CAR.  This is an unexpected result in the tunneling limit, where theory predicts that EC and CAR are equal and opposite in sign at all energies, so that no nonlocal voltage should be observed.\cite{falci} 

\begin{figure}
      \includegraphics[width = 8.5cm]{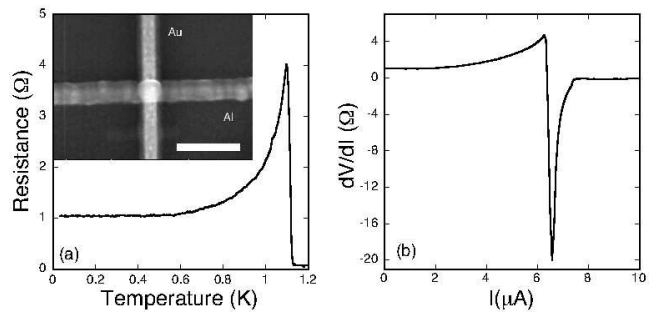}
      \caption{{\small \textbf{(a)} Temperature dependent four-terminal resistance of a NS junction shown in the inset.  For the measurement, a small ac current is sent from one Au lead to one Al lead and the resulting four-terminal resistance is measured with a bridge using a lock-in amplifier with the two remaining leads.  The size bar in the inset is 300 nm.  \textbf{(b)} Differential resistance $dV/dI$ as a function of the dc current $I$ through the junction measured at 300 mK.  $I$ is injected in parallel to the ac current for this measurement.\vspace{-0.60cm}
       } \label{fig2}}
\end{figure}

We report here measurements on NS devices with highly transparent interfaces, which we have found show large nonlocal voltages below the superconducting transition as a function of both temperature and applied current bias.  Two specific types of nonlocal signals are observed:  the first type is more prominent at very low temperatures and current bias and decays on a short length scale comparable to $\xi_S$.  The second type of nonlocal voltage is largest near $T_c$ and near the critical current and decays on a longer length scale.  Based on their dependence on temperature, current bias and distance, we can readily distinguish the two types of nonlocal signals, and identify the first type as arising from EC and CAR, and the second type as arising from CI.  The EC/CAR signal in our devices is large and positive, as predicted by theory for the case of transparent contacts, where the EC contribution is expected to be larger than CAR.\cite{melin}

The characteristic difference between the EC/CAR contributions and the CI contribution can be seen in the temperature-dependent resistance of the simple NS cross device shown in Fig. \ref{fig2}(a).  Above $T_c$, the normal state resistance $R_N$ of the interface is 70 m$\Omega$.  This should be compared to a rough estimate of the Sharvin resistance of the NS junction based on its dimensions, which is of the order of 10 m$\Omega$, indicating that the NS interface is highly transparent.  As the junction is cooled through $T_c$, a large resistance peak is observed.  This resistance peak, due primarily to the CI contribution, decreases as the junction is cooled further, but then saturates at around 1 $\Omega$, much larger than $R_N$.  As noted earlier, the CI contribution is expected to vanish as $T\rightarrow0$, hence the low-temperature resistance arises from EC and CAR.  Similar behavior is observed in the low-temperature differential resistance $dV/dI$ of the junction as a function of the dc current $I$, shown in Fig. 2(b).  The zero-bias differential resistance is approximately 1 $\Omega$, reflecting the low-temperature limit of the resistance seen in Fig. \ref{fig2}(a).  As $I$ is increased, a peak in $dV/dI$ develops, corresponding to the CI contribution of Fig. \ref{fig2}(a).  As $I$ is increased further, however, a large negative differential resistance is observed before the junction returns to its normal-state resistance at $I  \approx 6$ $\mu$A.  The differential resistance peak and dip are primarily associated with CI, although we cannot preclude the possibility that contributions from EC/CAR may be present as well.

\begin{figure}[!]
      \includegraphics[width=8.5cm]{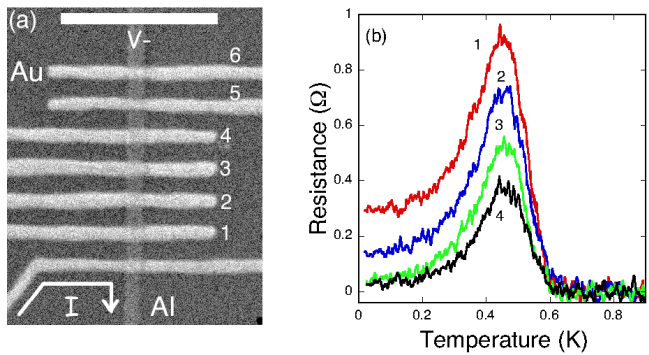}
      \caption{{\small (color online) \textbf{(a)} Scanning electron micrograph of a device with six nonlocal leads, each separated by 210 nm.  For the measurements, a current $I$ is sent through the lower Au lead into the superconductor as shown, and the resulting four-terminal nonlocal resistance is measured between the Au leads marked 1 through 6 and the superconducting lead labeled $V-$.  A small ac current is used for the temperature dependence measurements, and a dc current $I$ is injected in parallel to the ac current for the differential resistance measurements.  The size bar is 1 $\mu$m.  \textbf{(b)} Temperature dependence of the nonlocal resistance of the sample in (a) measured using the first four leads.  The numbers labeling the curves correspond to the lead numbers in (a).\vspace{-0.60cm}
       } \label{fig3}}
\end{figure}

In order to distinguish between EC/CAR effects and CI, we fabricated NS devices that enable us to measure the truly nonlocal differential resistance as a function of the distance from the NS interface.  Fig. \ref{fig3}(a) shows a scanning electron microscope (SEM) image of one of these devices. These samples were fabricated using multilevel e-beam lithography on oxidized Si substrates.  The 50 nm thick Au layer was patterned and deposited first. After further patterning for the Al layer, an \textit{in situ} Ar plasma etch was performed prior to the deposition of the 50 nm Al layer to ensure transparent interfaces.   Figure \ref{fig3}(b) shows the nonlocal resistance as a function of temperature measured using the four normal metal probes closest to the NS interface through which the current is sent.  The curves are similar to the data for the NS junction shown in Fig. \ref{fig2}(a).  In this case, however, the nonlocal resistance vanishes above $T_c \sim$ 0.6 K, as expected in the normal state.  (We note that $T_c$ for the superconducting wire in this device is suppressed from its typical value of $\sim$1.2 K for Al due to the inverse proximity effect arising from the large number of normal probes in contact with the superconductor.)  Below $T_c$, a peak develops; as the temperature is lowered further, the resistance saturates at a finite value.  Measurements of the differential resistance of the interface itself give an even larger low-temperature resistance similar to that seen in Fig. \ref{fig2}(a), indicating that the origin of the large zero-bias offsets seen in Figs. \ref{fig2}(a) and \ref{fig3}(b) is the same.  Both the magnitude of the resistance peak and the low-temperature nonlocal resistance decrease as one moves further away from the NS interface; however, the length scales over which they decay are different, with the length scale for the low-temperature resistance being shorter as can be seen from Fig. \ref{fig3}(b).

\begin{figure}[!]
      \includegraphics[width=8.5cm]{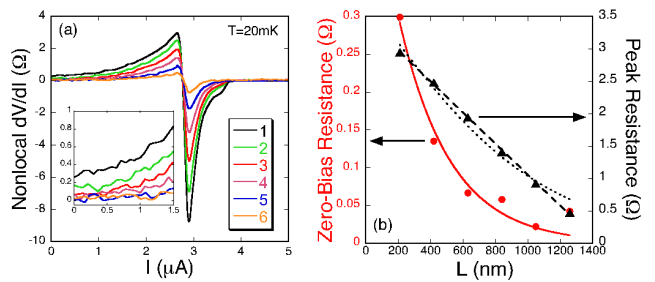}
      \caption{{\small (color online) \textbf{(a)} Nonlocal differential resistance of the device of Fig. \ref{fig3}(a) at 20 mK.  The numbers refer to the Au probes in Fig. \ref{fig3}(a) with which the signal was measured.  Inset:  expanded view of the low current bias regime, showing that the zero-bias differential resistance decays rapidly as one moves away from the NS interface through which the current flows.  \textbf{(b)} Zero-bias ($I$=0) resistance (circles) and  finite-bias resistance peak (triangles) from (a) as a function of the distance, $L$, of the nonlocal probe from the current-carrying NS interface.\vspace{-0.50cm}} \label{fig4}}
\end{figure}

Similar behavior is observed in the nonlocal differential resistance $dV/dI$ as a function of the dc current $I$ sent through the interface.  Figure \ref{fig4}(a) shows the resulting nonlocal differential resistance at a temperature of 20 mK for all six voltage leads shown in Fig. \ref{fig3}(a).  Due to a small asymmetry in $dV/dI$ vs. $I$ that likely arises from thermopower effects, these curves show only the symmetric component of $dV/dI$.  The curves are similar in form to the curve for the NS cross shown in Fig. \ref{fig2}(b), but show the decrease of the nonlocal differential resistance with distance from the NS interface.  While the finite-bias peak is observable even to the sixth lead, the inset to Fig. \ref{fig4}(a) shows that the zero-bias offset essentially vanishes by the third lead.
The different dependence of these two nonlocal contributions on the distance from the interface, $L$, can be seen more explicitly by plotting the magnitude of the zero bias differential resistance and the finite-bias peak resistance as a function of $L$.  These data are shown in Fig. \ref{fig4}(b).  The zero-bias resistance clearly decays on a much shorter length scale than the finite-bias peak resistance.  The theoretical decay of EC/CAR depends upon the geometry of the sample; for the samples measured here the NS interface is comparable in its dimensions to $\xi_S$, approaching the extended contact limit where an exponential decay is predicted.\cite{feinberg}  The solid line shows the best fit of the zero-bias resistance to an exponential decay,  $e^{-L/\xi_S}$, with a value of $\xi_S$= 315 nm.  A slightly better fit, yielding $\xi_S$= 217 nm may be found by adding a small, constant zero-bias resistance to the exponential decay.  Typical values for $\xi_S$ for the diffusive Al films with $T_c \sim$ 1.2 K deposited in our laboratory are 140-310 nm.\cite{chien}  The fitted values of $\xi_S$ are in very good agreement with these values given that $T_c$=0.6 K due to the inverse proximity effect. 

 The dotted line in Fig. \ref{fig4}(a) shows a fit of the peak resistance to $(1-\tanh(x/\Lambda_{Q^*}))$, the form expected if the Au leads probe the quasiparticle CI potential $\mu_{qp}(x)$, with a value of $\Lambda_{Q^*}$= 1085 nm.  We note that an even better fit is obtained with a linear decay, $(\Lambda_{Q^*} - x)$, with $\Lambda_{Q^*}$= 1442 nm.  We have no explanation for this observation at present, but we note that a linear decay is expected if the Au leads probe the normal-state potential $\mu_N(x)$ rather than the quasiparticle potential $\mu_{qp}(x)$.  The values of $\Lambda_{Q^*}$ obtained from the fits are shorter than the value of 10-20 $\mu$m typically quoted for Al from experiments on charge imbalance phenomena performed almost three decades ago.\cite{chi,stuivinga,mamin}  More recent experiments on charge imbalance are consistent with $\Lambda_{Q^*} \sim$ 1 $\mu$m, in agreement with the values we obtain.\cite{strunk,chien}

We can therefore distinguish the EC/CAR contribution to the nonlocal resistance from the CI contribution based its temperature, current and scaling with the distance from the current-carrying NS interface.  The EC/CAR contribution at low temperature and zero current-bias is positive, indicating that the magnitude of the EC contribution is larger than that of the CAR contribution, as predicted for transparent NS interfaces.\cite{melin}   We cannot preclude the fact that there may also be contributions due to EC and CAR at finite bias.  Indeed, there are experimental indications that there are contributions other than those due to CI in the current regime where the resistance peak and dip are observed in the differential resistance. If we scale the the $dV/dI$ curves so that the resistance peaks for all curves lie on top of one another, the resistance dips will not similarly scale, as would be expected if both resistance peaks and dips were due to CI.  Thus, it is quite likely that there may be contributions due to EC and CAR at finite current bias.  These contributions are not precluded by the theory, although the detailed dependence on temperature and current bias in the regime of transparent interfaces is still to be worked out.

We thank D. Feinberg and D. Beckmann for useful discussions. This work was supported by the NSF through grant DMR-0604601.

\end{document}